# View-Centric Modeling of Automotive Logical Architectures

Hans Grönniger[1], Jochen Hartmann[2], Holger Krahn[1],
Stefan Kriebel[2], Lutz Rothhardt[2], and Bernhard Rumpe[1]

[1]Institut für Software Systems Engineering, TU Braunschweig, Germany
[2]BMW Group, München, Germany

**Abstract:** Modeling the logical architecture is an often underestimated development step to gain an early insight into the fundamental functional properties of an automotive system. An architectural description supports developers in making design decisions for further development steps like the refinement towards a software architecture or the partition of logical functions on ECUs and buses. However, due to the large size and complexity of the system and hence the logical architecture, a good notation, method, and tooling is necessary. In this paper, we show how the logical architectures can be modeled succinctly as function nets using a SysML-based notation. The usefulness for developers is increased by comprehensible views on the complete model to describe automotive features in a self-contained way including their variants, modes, and related scenarios.

## 1 Introduction

Developing automotive embedded systems is becoming a more and more complex task, as the involved essential complexity steadily increases. This increase and the demand for shorter development cycles enforces the reuse of artifacts from former development cycles paired with the integration of new or changed features.

The AUTOSAR methodology [AUT] aims to achieve reuse by a loosely coupled component-based architecture with well-defined interfaces and standardized forms of interaction. The approach targets at the software architecture level where the interface descriptions are already detailed and rich enough to derive a skeleton implementation (semi-)automatically. Despite its advantages software architectures are less useful in early design phases where a good overview and understanding of the (logical) functions is essential for an efficient adaptation of the system to new requirements. Detailed interface descriptions and technical aspects would overload such a description. In compliance with [Gie08] we argue that the component-based approach has its difficulties when applied to automotive features that highly depend on the cooperation of different functions. To comprehend the functionality of features, it is essential to understand how functions interact to realize the feature. Complete component-like descriptions of single functions are less useful as they only describe a part of a feature while a comprehensive understanding is necessary.

We identified the following problems with notations and tools proposed or used today:

- Tools that only provide full views of the system do not scale to a large amount of functions.

- Notations that have their roots in computer science are not likely to be accepted by users with a different professional background.
- Under-specification and the abstraction from technical details are often excluded if tools aim at full code generation.

Our contribution to the problem of modeling complex embedded automotive systems is thus guided by the following goals:

- A comprehensible description of functions, their structure, behavior, and interactions with other functions should be supported.
- Interesting functional interrelations should be presentable in a way such that they are comprehensible for all developers involved.
- The system shall be modeled from different viewpoints in a way that a developer can concentrate on one aspect at a time and the conformance to the complete system can be checked automatically. In addition, the developer shall be supported in understanding how a certain viewpoint is realized in the complete system.

In accordance to [vdB04, vdB06], we argue that function nets are a suitable notation for describing the logical architecture of an automotive system. We further explain how views for features including their modes and variants can help to ease the transition from requirements to a logical architecture. In addition to the aspects already explained in our previous work [GHK+07, GHK+08], we extend the view-based approach to model scenarios with UML communication diagrams [OMG05] that are also consistent to the logical architecture or other views.

The rest of the paper is structured as follows. In Section 2 function net architectures and views are explained. Section 3 describes how this notation can be used within an automotive development process. Section 4 focuses on scenarios which complement the feature views to simplify the understanding and enhance the testability. Section 5 presents related work and Section 6 concludes the paper.

## 2 Function Nets Architecture and Views

For modeling the logical architecture as function nets an appropriate notation has to be found. We evaluated UML 2.0 [OMG05] and other UML derivates like UML-RT [SGW94] and SysML [OMG06] which in general were found suitable for architecture and function net modeling [RS01, vdB04]. We favored SysML over other notations because it uses Systems Engineering terminology which is more intuitive for people with different professional backgrounds than others which have their roots in computer science. SysML internal block diagrams allow us in contrast to UML composite structure diagrams to model cross-hierarchy communication without using port delegation which was helpful to model static architectures. A more detailed discussion can be found in [GHK+07].

We use a subset of SysML internal block diagrams to enable compact definitions and decrease the learning effort for the notation. An internal block diagram (ibd) used as a function net may only contain directed connectors to indicate the signal flow direction. Multiplicities of blocks and connectors have always the value one and are therefore omitted. An example of a function net diagram can be found in Figure 1. It shows a simplified complete function net "CarComfort" which contains a fictitious model of a central locking functionality. It evaluates the driver's request and opens or closes the doors accordingly. In addition, the doors close if the vehicle exceeds a certain speed limit (auto lock).

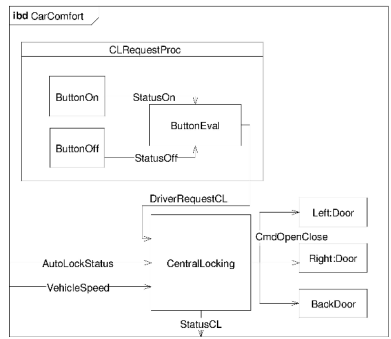

Figure 1: Part of an automotive function net

Syntactically, the function net is a valid SysML internal block diagram. In that example, we used three layers of hierarchy (top-level, block "CLRequestProc" and, e.g., "ButtonOn"). With the signal "DriverRequestCL", it also shows an example of cross-hierarchy communication. Instantiation is also supported. In the example, there are two doors which are instantiated by giving each block a name, in this case "left" and "right". These two blocks share their behavior but not their internal state. Instantiation is an important feature that allows one to reuse of a block multiple times which also avoids redundant block definitions that are poorly maintainable [GHK+07].

Views are also modeled as internal block diagrams indicated by the specific diagram use ≪view≫. In general, views focus on a certain aspect of the function net. By adding a few specific properties, views can also model the environment and context of the considered aspect. Compared to the complete function net, a view may leave out blocks, signals, or hierarchy information which is present in the related complete function net. "Env(ironmental)" blocks and non-signal communication can be added to increase the understandability. The environmental elements refer to non E/E elements that have a physical counterpart. Non-signal communication is modeled by connectors that are connected to special ports in which "M" represents mechanical influence, "H" hydraulics, and "E" electrical interactions. Blocks marked with the stereotype "ext(ernal)" are not central to the view but are necessary to form an understandable model.

The example in Figure 2 shows that whole blocks from the complete function net have been left out (block "CLRequestProc") and that the blocks "CentralSettingsUnit" and "VehicleState" have been included in the view to clarify where the signals "AutoLockStatus" and "VehicleSpeed" originate. The physical door look "LockActuator" is shown in the figure as an environmental element (marked with the corresponding stereotype ≪env≫).

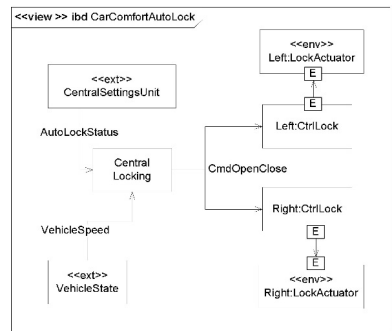

Figure 2: View of the autolock functionality including external blocks and environment

Like explained in [GHK+07, GHK+08], a view and a complete function net of the system are consistent if the following consistency conditions hold.

1. Each block in a view not marked with the stereotype ≪env≫ must be part of the logical architecture.

2. Views must respect whole-part-relationships. If there is such a relationship between functions in the view, it must also be present in the complete function net. Intermediate layers may be left out, though.

3. The other way round, if two functions are in a (possibly transitive) whole-part-relationship in the complete function net and both are present in a view, they must have this relationssship in the view, too.

4. Normal communication relationships (except those originating from M, E, or H ports) shown in a view must be present in the logical architecture. If the view indicates that certain signals are involved in a communication they must be stated in the architecture. If no signal is attached to a communication link in a view at least one signal must be present in the architecture.

5. A communication relationship needs not be drawn to the exact source or target, also any superblock is sufficient if the exact source or target is omitted in the view.

One view on a function net is a specialization of another view if both are consistent to the same complete function net and the following context condition holds.

6. The blocks and connectors in the specialization function net are a subset of the blocks shown in the referred function net.

## 3  Function Nets and Views in the Automotive Development Process

The above described mechanisms to model the logical architecture as function nets and to create views on the complete model can be used for a variety of purposes. Mainly the views can be used to model an automotive feature in a self-contained way. The term *feature* denotes a functionality that is perceptible by customers (e.g., a braking system) or an automotive engineer (e.g., system functions that effect the whole vehicle like diagnostics). Further views which are usually specializations of a feature view are used to model the variants and modes of that feature. The variants provide the same principle functionality but are distinguishable by having different individual properties. The complete function net includes all variants and through parameterization the intended variants are chosen. Modes are used to show that features change their observable behavior in a distinct way when certain conditions are fulfilled. Modes are described by a statechart where the state names correspond to the different modes of operation, whereas the transitions are used to describe the conditions which force a change of the mode of operation. As error degradation is a common example for using modes, views are then used to show the active subsystems and leave the inactive subsystems out. Further details about the use of views for modeling features, variants, and modes can be found in [GHK$^+$08, GKPR08].

From a technical position, a view is consistent with a function net or view if the given constraints hold. Especially, each developer decides on his/her own which external or environmental elements are necessary to form a self-contained description. In a company where many developers concurrently work on a solution, this raises the question for modeling guidelines which standardize the use of additional elements and therefore simplify cooperative development.

Figure 3 gives an overview of the development process. The requirements are captured textually in a distributed manner and form the basis for modeling features in a self-contained way by using feature and other according views. These models ease the transition to the logical architecture mainly because the same notation is used and automatic checks for conformance can be applied on the basis of the above explained consistency conditions. The logical architecture can then be further refined towards an AUTOSAR software and hardware architecture. The realization is then derived within the AUTOSAR process.

Please note that the figure concentrates on the structural view point only. This view point forms the backbone of automotive software development, but has to be complemented by behavioral descriptions on the one hand, and exemplary scenarios on the other hand. For each of the view points, different diagrams exist that are related to the structural diagram on each layer of abstraction.

The feature models themselves are a unit of reuse: When moving from one product line to another the still relevant features of the old product line are determined and the necessary changes on requirements are made. The according changes are reflected in the features views. Finally, the traceable connection between the feature views and the logical architecture helps to define the relevant changes in the complete function net. During the transition from one development step to another different design decisions are made. For example, going from the logical architecture to the technical architecture the developer has to determine the physical location of a logical function on a certain ECU. The reuse of existing software from a former product line is often only possible if the function is not mapped to another ECU with completely different properties. Therefore, annotations can be added to the logical architecture resp. a feature view to pre-determine the result of the placement decision. This restriction on the design space maximizes reuse.

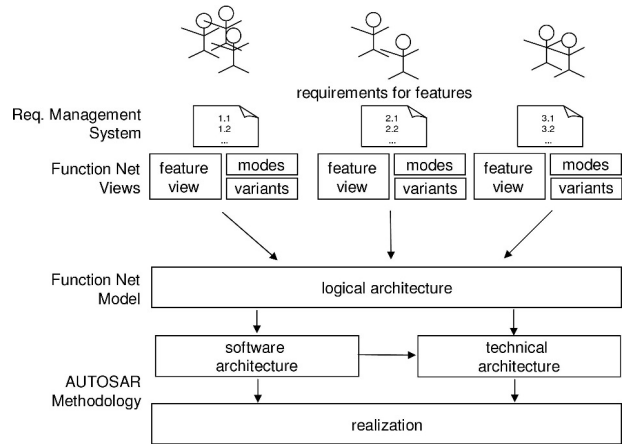

Figure 3: Automotive development process

## 4 Using Views to Describe Scenarios

As described above views can be used to model features in a self-contained way. Further specializations on this view can be used to explain the according modes and variants. This form of specification is supplemented by behavioral descriptions of the non-composed blocks using an arbitrary form of description like statecharts, Matlab/Simulink models or plain text depending on the chosen abstraction layer. These models aim at a complete description of the considered system or feature resp. their modes and variants. Nevertheless, sometimes (e.g., for safety assessments) it is important to document how the system reacts to certain external events or failures of subsystems in an exemplary fashion. In addition, the scenarios help developers to understand the system better by showing representative situations. For this kind of exemplary behavior we adopted the UML communication diagram [OMG05]. In this kind of diagram the subsystems exchange signals in the enumerated order. The basic form uses another view on the corresponding complete, feature, mode, or variant function net which includes only the active subsystems in the scenario. Therefore, communication diagrams can be used on both abstraction layers, on the feature level as scenarios on feature views and on the complete function net where scenarios for a composed automotive system are modeled.

The modeled system may communicate by discrete events or by continuous signal flow. Therefore, we extended the form of communication normally used in communication diagrams by allowing the following conditions which help to describe continuous forms of interaction. The notation allows to specify multiple ranges for a certain signal, e.g., to specify that a signal is either invalid or greater than a certain value.

- The value of a signal s is larger than v: $s > v$

- The value of a signal s becomes larger than v: $s >> v$

- The value of a signal s has a certain value v : s == v
- The value of a signal s changes to a certain value v : s = v
- The value of a signal s is smaller than v: s < v
- The value of a signal s becomes smaller than v: s << v

For discrete signals also the following notation is useful:

- The value of a signal s changes from v to w: s : v −> w

An example for a scenario that is modeled by a communication diagram can be found in Figure 4 with a corresponding variant view of the CentralLocking function. More details on the variant mechanism used can be found in [GHK+08, GKPR08]. In the example the signal "VehicleSpeed" exceeds the value of 10km/h. Therefore, the block "Eval-Speed" changes its output value from "Open" to "Close". After that, the output value of the block "Arbiter" must be "Close" and therefore the doors of the car will close (not modeled here). This interaction is modeled as `CmdOpenClose == Close` and not as `CmdOpenClose : Open -> Close` because the Arbiter may already have closed the doors by sending the output signal value "Close", e.g., because of input values from the not modeled input signal "AutoLockStatus".

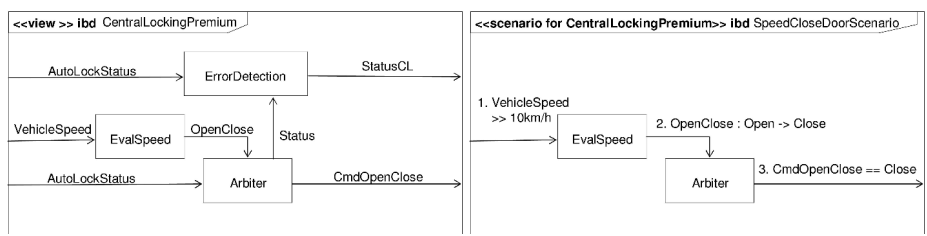

Figure 4: Variant view and a corresponding scenario

Scenarios like described above could also be modeled using sequence diagrams. Discussions with developers turned up that the similarity between the complete function net and the scenarios are helpful to understand the considered system. However, sequence diagrams might be helpful to model complex interactions between few communication partners, whereas simple interactions between many communication partners can be modeled more concisely in a communication diagram. Therefore, we allow using both, sequence and communication diagrams, for modeling scenarios.

Nevertheless, this form of communication diagrams can easily be transformed to a sequence diagram with the same meaning. Therefore, extensions of sequence diagrams like explained in [Rum04] could also be used in communication diagrams. The main new concepts are the following:

- Communications can be marked by the stereotype ≪trigger≫ to mark that they trigger the scenario.

- The matching policy of the communication partners can be marked as either
    - *complete* to indicate that all occurring communication is already shown in the diagram. All additional communication that is observed in a system run is interpreted as wrong behavior.
    - *visible* to indicate that all communication between blocks shown in a diagram is actually modeled. However, there may exist other blocks not shown in the scenario that also communicate in a system run.
    - *free* to indicate that arbitrary interactions may occur additionally to the explicitly modeled communications.

As described in [Rum04], these extensions can be used to automatically derive a test case from the communication diagram (with the sequence diagram as an intermediate step). The interactions in the sequence diagram are assumed to be ordered as they appear in the diagram. Additional invariants could be used to model timing constraints in the scenarios. The test cases use the triggers to invoke the actual implementation and an automaton that checks if the system shows the correct behavior. This method can be applied only if there is a traceable connection from the function nets to the actual realization. The connection is used to translate the triggers which are logical signals to, e.g., messages in the realization. Vice versa, the observed messages in the realization are transformed back to the according logical signal flow that is checked against the described interactions. If such an approach is technically to difficult, the scenarios can still be used for validating and testing the logical architecture in a simulation.

## 5 Related Work

Function net modeling with the UML-RT is described in [vdB04]. We extended this approach by using the SysML for modeling function nets and explained its advantages. We supplement the approach by views that simplify the transition from requirements to logical architectures in early design phases as they are able to model features and their variants, modes, and also scenarios.

In [RFH+05, WFH+06] service oriented modeling of automotive systems is explained. The service layer is similar to the modeling of features. Another example of an architecture description language that aims at supporting a development process from vehicle requirements to realization is being defined in the ATESST project [ATE] based on EAST-ADL [EAS]. Both works do not explicitly provide support for modeling exemplary scenarios. In addition, we explored how feature descriptions can benefit from modeling the environment together with the feature. AML [vdBBRS02, vdBBFR03] considers abstraction levels from scenarios to implementation. Unlike our use of scenarios, scenarios in [vdBBRS02] are employed in a top-down development process on a higher level of abstraction out of which the logical architecture will later be developed.

In [DVM+05] the use of rich components is explained that employ a complex interface description including non-functional characteristics. In that approach, scenarios could be modeled as viewpoints of a component. In contrast to our approach rich components focus less on the seamless transition from requirements to function nets but assume an established predefined partitioning in components.

The AUTOSAR consortium [AUT] standardizes the software architecture of automotive system and allows the development of interchangeable software components. One main problem of this approach is that software architectures are too detailed in early development phases where functions nets are commonly accepted by developers. In addition, like also argued in [Gie08], the component-based approach has its difficulties when applied to automotive features that highly depend on the cooperation of different functions.

# 6 Conclusion

In this paper, we summarized our approach for modeling the logical architecture of automotive systems using views. We regard views as suitable for describing automotive features, but also their variants and modes. More information can be found in [GHK$^+$07, GHK$^+$08]. In this paper we additionally introduced a scenario notation that is similar to UML communication diagrams but is a view on a function net at the same time. The syntactic proximity to the logical architecture simplifies the understanding of scenarios. Scenarios can be used to document important use-cases of features. These use-cases contain information that help developers to understand and correctly implement the intended functionality.

Despite the scenarios which show exemplary behavior, the approach described in this paper focuses on structural aspects only. This is mostly sufficient for the modeling of a logical architecture because architectures mainly focus on structural properties and decomposition of functionality. For the self-contained modeling of features this approach has to be extended with full behavioral specifications to gain its full usefulness.

When the proposed model types (complete function nets, views for features, modes, variants, and the described scenarios) are used to model an automotive system, a greater number of models have to be created and references between models have to be establised. We plan to investigate into existing model mangement strategies and adopt them to the needs of a model-based automotive development process in order to handle the large number models.